\documentclass[]{aa}  
\usepackage{natbib}
\bibpunct{(}{)}{;}{a}{}{,} 

\usepackage{graphicx}
\usepackage{txfonts}
\usepackage{hyperref}
\usepackage[svgnames]{xcolor}
\newcommand{\teff}  {T$_\mathrm{eff}$}

\begin{document}

\title{Exoplanet hosts reveal lithium depletion}

    \subtitle{Results from a homogeneous statistical analysis}

\author{P. Figueira\inst{1},
       \,J. P. Faria\inst{1,2},
	\,E. Delgado-Mena\inst{1},
	\,V. Zh. Adibekyan\inst{1},
	\,S. G. Sousa\inst{1,2},
	\,N. C. Santos\inst{1,2},
	\and
	\,G. Israelian\inst{3,4}
         }

   \institute{Centro de Astrof\'{i}sica, Universidade do Porto, Rua das Estrelas, 4150-762 Porto, Portugal\\
     \email{pedro.figueira@astro.up.pt}
	\and
    Departamento de F\'{i}sica e Astronomia, Faculdade de Ci\^{e}ncias, Universidade do Porto, Portugal
    \and
    Instituto de Astrof\'{\i}sica de Canarias, C/ Via Lactea, s/n, 38200, La Laguna, Tenerife, Spain
    \and
    Departamento de Astrof\'isica, Universidad de La Laguna, 38205 La Laguna, Tenerife, Spain }

   \date{}

 
  \abstract
   {}
   {We study the impact of the presence of planets on the lithium abundance of host stars and evaluate the previous claim that planet hosts exhibit lithium depletion when compared to their non-host counterparts.}
   {Using previously published lithium abundances, we remove the confounding effect of the different fundamental stellar parameters by applying a multivariable regression on our dataset. In doing so, we explicitly make an assumption made implicitly by different authors: that lithium abundance depends linearly on fundamental stellar parameters. Using a moderator variable to distinguish stars with planets from those without, we evaluate the existence of an offset in lithium abundances between the two groups. We perform this analysis first for stars that present a clear lithium detection exclusively and include in a second analysis upper lithium measurements.}
   {Our analysis shows that under the above-mentioned assumption of linearity, an offset in lithium abundance between planet hosts and non-hosts is recovered. This offset is negative, showing an enhanced depletion for planetary hosts, and is a statistically significant result. By bootstrapping the error bars, we concluded that an inflation on the lithium uncertainty estimations by a factor of larger than 5 is required to render the measured offset compatible with zero at less than 3-4\,$\sigma$ and make it non-significant. We demonstrated that the offset as delivered by our method depends on the different nature of the stars in the two samples. We did so by showing that the offset is reduced down to zero if the planet-host stars are replaced by comparison stars in a mock planet-host sample. The offset is also shown to be significant at 3.75\,$\sigma$ when compared with that of a population in which planet-host and comparison tags are shuffled, representing a situation in which the tagging is decorrelated from the presence of orbiting planets. Moreover, the measured depletion is still significant when one imposes different constraints on the dataset, such as a limit in planetary mass or constrain the host temperature to around solar value. We conclude then that planet-host stars exhibit enhanced lithium depletion when compared with non-host stars.}
   {}

  \authorrunning{P. Figueira et al.}

  \maketitle
  \titlerunning{\thetitle}
%

\section{Introduction}\label{sec:Intro}
Light elements, such as lithium (Li), are good tracers of stellar internal mixing and rotation \citep[e.g.][]{pinsonneault90,stephens97}. This element is easily destroyed in the inner layers of solar-type stars, mainly during the pre-main sequence, but its destruction can also take place in stellar envelopes if an efficient mixing process is at work. Therefore, the study of Li abundances may be key to understanding processes related to the angular momentum evolution of planet-host stars and, as such, the mechanisms behind the formation of planets.

The work of \citet{king97} was the first to suggest a connection between Li depletion and planet hosts after finding a difference in abundance for the very similar stars of the double system 16\,Cyg, one of them hosting a Jupiter. From then on, the Li dependence on the presence of planets has been extensively discussed in the literature. On one hand, several independent groups find that planet hosts with \teff\ close to the solar value exhibit lower abundances of Li when compared to non-hosts \citep[e.g.][]{israelian04,chen06,gonzalez_li08,israelian09,takeda10,gonzalez_li10,sousa_li}. 
On the other hand, several other works report no dependence \citep{ryan00,luck06,baumann,ghezzi_li,ramirez_li12} and claim that the difference in Li abundances between both populations is produced by a bias in the [Fe/H] and age distribution of the samples. More metallic stars are expected to have thicker envelopes and thus stronger depletion, and lower Li abundances can be explained if the population of planet hosts is older than the non-hosts. In a nutshell, the abundance of Li is known to depend on several fundamental stellar parameters, chiefly among them \teff, [Fe/H], log\,$g$, and Age, but the functional form of this intricate dependence has not been pinpointed yet. Since each fundamental parameter varies across a wide range of values from one star to the other, it is very hard to isolate the influence of each one of these factors\footnote{Moreover, it is well known that some of these parameters are strongly correlated, a point that makes the analysis and interpretation extremely challenging.}.

The discussion on Li depletion on exoplanet hosts was revived by the recent works of \cite[][--\,DM14]{2014A&A...562A..92D} and \cite[][--\,G14]{2014MNRAS.441.1201G}, who concluded that exoplanet hosts show an appreciable Li depletion when compared with non-planet hosts. The impact of the fundamental parameters on Li determination was addressed in two radically different ways. DM14 constrained the impact of such parameters by studying solar twins and restricting the study sample to a set of stars with narrow parameter variations around solar values. G14 followed a different approach and defined an index to estimate the similarity between stars with planets and comparison stars, 

\vskip-1.5em
\begin{eqnarray}\label{Gonzalez:Delta}
 \Delta_{p,c} & = &30 \,|\mathrm{log T^c_{eff} - log T^p_{eff}}|\,+ \, |[Fe/H]^c - [Fe/H]^p| \\ \nonumber
              && +\, 0.5\,|\mathrm{log}g^c - \mathrm{log}g^p| \,+\, |\mathrm{log\, Age}^c - \mathrm{log\, Age}^p| ,
\end{eqnarray}

where the upper-script ``p'' refers to planet hosts and the upper-script ``c'' refers to comparison stars. This index was used to weigh (using as weight 1/$\Delta_{p,c}^2$) the differences in Li abundance, allowing for a meaningful comparison across a wide range of stellar parameters. The lower the value of the index, the higher the degree of similarity between the stars; in this formula, the coefficients of the different parameters reflect the expected impact of a variation of that parameter on the final Li abundance. 
This approach is not entirely new: \cite{1994A&A...287..191P} were the first to explore the dependence of Li abundance on several fundamental stellar parameters. The authors made use of a multiparameter fit to evaluate the impact of each parameter on Li abundance. The same approach was recovered by \cite{2000AJ....119..390G}, who increased the sample size and searched for differences between planet hosts and a comparison sample. 

The problem of the dependence of Li abundance on stellar parameters can be looked upon from a more general perspective. From the statistical point of view, removing the effect of an independent variable, such as a stellar parameter, on a dependent variable, such as the Li abundance, is called controlling for the variable. This allows us to reduce the confounding effect of the controlled variables and measure the effect of the remaining (non-controlled) variables on the dependent variable. Our problem is a particular case of such an analysis, as we want to test whether two groups, a control group (the comparison stars) and the study group (the planet hosts) exhibit an appreciable difference in Li content. To do so, we consider the presence of what is called a moderator variable, a variable that is used to distinguish between two groups in the context of a global analysis.

Motivated by the recent work of G14, we used the aforementioned statistical concepts to address this long-standing question. We start by describing the dataset used in the analysis in Sect.\,\ref{sec:TheData}. In Sect.\,\ref{sec:MRA}, we describe the application of multivariable regression analysis to the Li measurements, and in Sect.\,\ref{sec:Censored}, we apply a generalization of the Tobit model to perform a censored regression on the the full dataset, including Li upper limits.  We discuss the results in Sect.\,\ref{sec:Discussion} and conclude in Sect.\,\ref{sec:Conclusions}\,.


\section{Data}\label{sec:TheData}

We analyze the sample studied in DM14. The values of \teff, [Fe/H], log\,$g$, and age, along with Li abundances and estimated associated uncertainty, are presented in Tables 3, 4, and 6 of this work. 
We did not take into consideration the few stars with no age determination. We note here that the Li uncertainty reported in DM14 (and used throughout the paper) mainly reflect the uncertainty in the determination of the continuum position but do not consider the effect of the absolute errors of stellar parameters on the Li abundance. To study the impact of an error on the stellar parameter determination, we changed the atmospheric parameters used in the atmospheric model (see DM14 for details) by 1\,$\sigma$ and with a relative sign, such as to maximize their impact on the Li uncertainty. The vast majority of the stars displayed a  small relative increase in uncertainty (below 10-20\,\%), which led us to keep the published values of DM14 as the baseline for our work. However, we stress that the reported uncertainties correspond to intrinsic errors and assume the stellar parameters are correct, and are thus underestimated with respect to the real ones.

For some of the stars studied, Li abundance was too low to be measured robustly, and only upper limits were obtained. When considering only the Li abundance determinations (Sect.\,\ref{sec:MRA}), our study encompassed 44 planet-host stars and 139 comparison stars, while when including the upper limits (Sect.\,\ref{sec:Censored}), the analysis was extended to a total of 90 planet hosts and 232 comparison stars.
We also considered a subset of our measurements, composed of only Jupiter-mass planets, with M\,$>$\,0.1$M_{\mathrm{Jup}}$, as defined in DM14, for analysis. This dataset was composed of 36 planet-hosts with Li abundance determinations and 79 planet-hosts with Li upper limits. For more details on the planet-host and non-planet-host samples, we refer the reader to DM14, in which the star's parameters are presented in detail.

For this study, it is fundamental to understand to which extent the planetary population orbiting around these stars was fully characterized, and their stars identified as planetary hosts. This is a challenging (and ongoing) task, but DM14 made use of several long-term surveys undertaken with the most precise spectrograph for radial-velocity measurements, HARPS \citep{2003Msngr.114...20M}, to minimize this issue. The dataset employed here is composed essentialy by FGK stars observed in the context of the HARPS GTO programs, for which the detection limits have been thoroughly analyzed \citep{2011arXiv1109.2497M}. From this analysis, we can conclude that the detectability of planets is overall high and particularly so for the Jupiter-mass planets; for this mass domain, the detectability level is close to 100\% for periods of up to several years, and we can then consider the planet-host identification to be complete and trustworthy for at least this type of planets.

\section{Multivariable regression analysis}\label{sec:MRA}

The simplest way to control an independent variable and thus reduce its confounding effect on a dependent variable is to perform a linear regression. Since we aim at controlling several variables simultaneously, the linear regression corresponds to a multivariable linear regression. We establish as working hypothesis that, Li abundance depends linearly on log(\teff), [Fe/H], log\,$g$, and log(Age) on first approximation, an hypothesis made explicitly by G14 through Eq.\,\ref{Gonzalez:Delta}\,, which in turn is inherited from previous works, such as \cite{1994A&A...287..191P} or \cite{2000AJ....119..390G}. However, we generalize upon it, as we do not assume an {\it a priori} value for the multiplying coefficients for each parameter.

A straightforward application of the regression analysis to each of the datasets can be performed. The result of the linear regression is then a function of the type 
\vskip-1.5em
\begin{eqnarray}\label{reg_basic}
\mathrm{log(A(Li))} & = & int. \,+\, \beta_1 \mathrm{log(}T_{eff}\mathrm{)} \,+\, \beta_2 [Fe/H] \,+ \, \beta_3 \mathrm{log}\,g  \\ \nonumber
                       && + \, \beta_4 \mathrm{log(Age)}
\end{eqnarray}

in which $int.$ represents the intercept value, which is the value of the dependent variable log(A(Li)) when all the independent variables are zero, and $\beta_{1-4}$ are the coefficients of the linear regression associated with each independent variable. The resulting coefficients from an ordinary least squares regression first on the planet-host and then on the comparison samples are provided in the first and second lines of the Table\,\ref{Table:MRAResults}, respectively. Several parameters present different values, but this is hardly surprising and provides little insight. By analyzing the two groups separately, there is no enforcing on the point that the dependence of A(Li) on the independent variables is the same for the two groups. For this fact to be taken into account, the regression has to be made on the two datasets simultaneously, while considering a variable to take the possible offset between the two groups into account. Such a dependence can be included on Eq.\,\ref{reg_basic} and takes the form

\begin{figure}

\includegraphics[width=9cm]{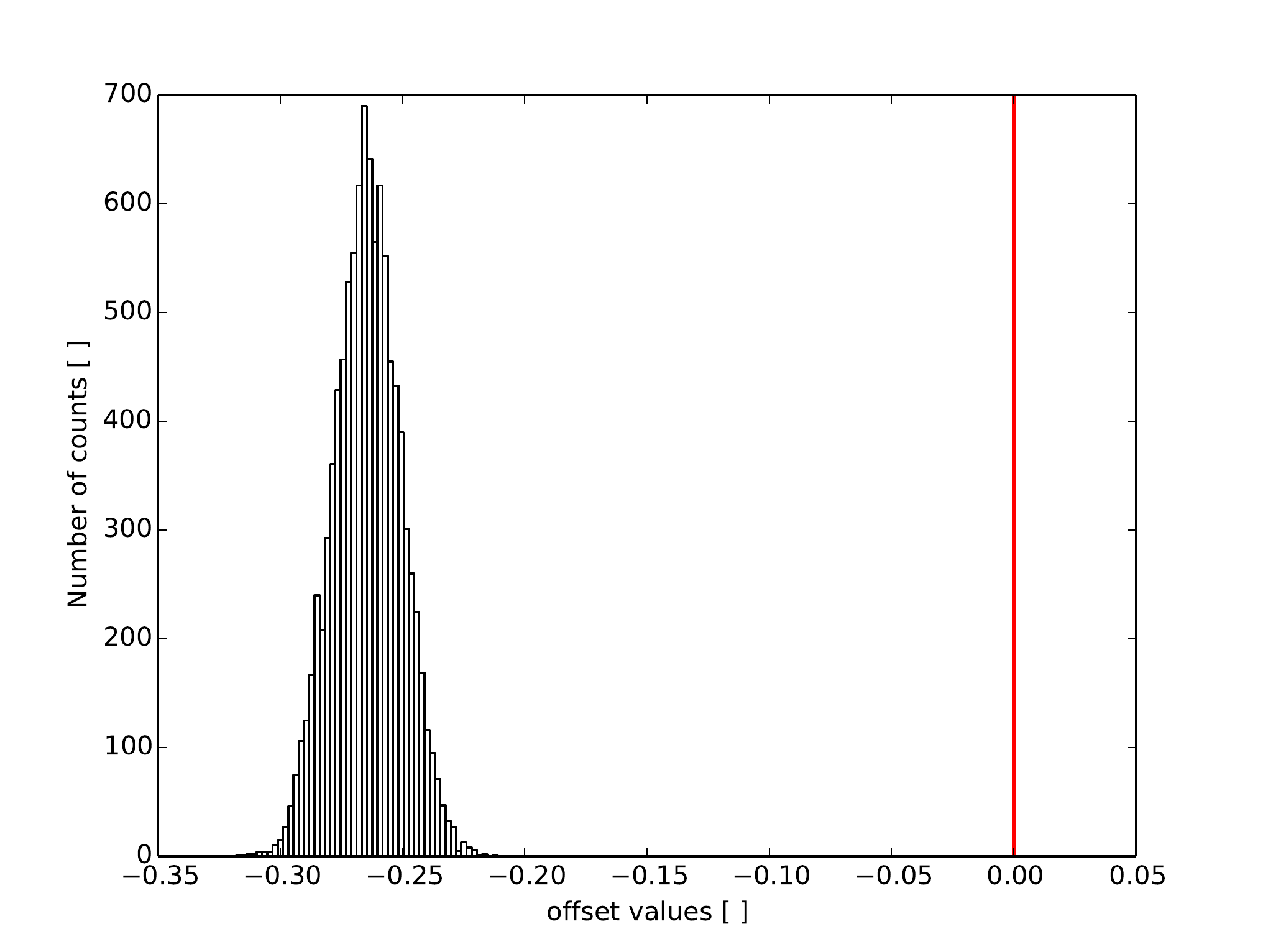}

\caption{The distribution of offsets for Eq.\,\ref{reg_moderated} as obtained by bootstrapping the Li values around the respective error bars. The vertical red line represents the zero value for the offset.  }\label{hist_off}

\end{figure}

\vskip-1.5em
\begin{eqnarray}\label{reg_moderated}
\mathrm{log(A(Li))} & = & int. \,+\, \beta_1 \mathrm{log(}T_{eff}\mathrm{)} \,+\, \beta_2 [Fe/H] \,+\, \beta_3 \mathrm{log}\,g \,+ \\ \nonumber
                       && + \, \beta_4 \mathrm{log(Age)} \,+\, M\,\times\,\mathrm{offset} ,
\end{eqnarray}

where the previously mentioned variables retain their meaning and two new variables are introduced: $M$ and offset. The variable $M$ is our categorical moderation variable (also known simply as ``dummy variable'') that takes as value 0 or 1; the former value is taken for the control group (i.e. comparison stars), and the latter for the study group (i.e. planet-host stars)\footnote{ Formally, a categorical moderation variable is an interaction variable that takes one of (at least) two values to indicate the absence or presence of some categorical effect and evaluate its impact on the dependent variable.}.
By doing so, one can apply the regression simultaneously on the two groups, obtaining a single value for $int.$ and $\beta_{1-4}$, and calculate the offset to apply to the study group. The regression on an equation of this form delivers the results presented on the third line of Table\,\ref{Table:MRAResults}. The offset is different from zero and much larger than the average error bar on Li abundance measurement, of around 0.07 (DM14), and thus expected to be meaningful, but a finer analysis is warranted. To estimate the significance of our result, we did a bootstrapping in which we applied a Gaussian distribution of scatter dictated by the corresponding uncertainty to each Li abundance. We then repeated the analysis, performing the regression on the Li values generated in this way a total of 10\,000 times. The results are presented in Fig\,\ref{hist_off}\,.

From the offset distribution, one can calculate that the observed value of -0.26 for the original dataset is at more than 19$\sigma$ from 0.0, and assuming a one-sided Gaussian distribution, this corresponds to a probability smaller than $10^{-4}$\% that this value is obtained by chance. This result is naturally heavily dependent on the error bars, and we thus proceeded to estimate the significance when each error bar is multiplied by an artificial factor $\alpha$. For $\alpha$\,=\,2.0, the significance drops to 9.6$\sigma$ and for $\alpha$\,=\,5.0 to 3.8$\sigma$. We can then conclude that an underestimation of the error bars on Li abundance relative to their real value by a factor of up to 5 does not prevent our result from being significant.

We repeated the analysis selecting only Jupiter-mass planets and present the results on the last line of Table\,\ref{Table:MRAResults}\,. The significance of the results is of 18$\sigma$, a value that drops to 9.0$\sigma$ and 3.6$\sigma$ when an $\alpha$ of 2 and 5 is applied, respectively. The difference in offsets resulting from the analysis for the two datasets is small, of 0.02 (and thus at 1.5$\sigma$), with the Jupiter-mass only dataset yielding the largest offset.

\begin{table*}

\caption{The parameters for each coefficient from multivariable linear regression analysis.} \label{Table:MRAResults}

\centering
\begin{tabular}{ccccccc} \hline\hline
 \ \ dataset &  {\it int.} &$\beta_1$ (log(\teff))& $\beta_2$ ([Fe/H]) & $\beta_3$ (log\,$g$) & $\beta_4$ (log(Age)) & offset \\ 
 \hline
planet hosts sample & -163.24 & 45.49 & -0.69 & -1.37 & -0.63 & ---  \\
comparison sample &  -73.49 & 21.27 & -0.25 & -0.99  & -0.67 &  ---  \\
\hline
combined sample &  -96.73 & 27.58 & -0.38 & -1.10 & -0.70 & -0.26  \\
combined sample, Jupiter-mass only & -92.21 & 26.42 & -0.37 & -1.14 & -0.71 & -0.28 \\

\hline 
\end{tabular}

\end{table*}

\section{Censored data analysis }\label{sec:Censored}

As discussed in Sect.\,\ref{sec:TheData}, for a significant fraction of the stars in our sample only upper limits for the Li abundance could be determined (for more details refer to DM14).
These values cannot be included in a straightforward (multivariable) regression analysis, and performing a constrained fit would neglect the true nature of the measurement, because the measurements are, in a sense, the bounded quantities. To correctly include the measured upper limits in the analysis, we used a censored regression model in which the dependent variable has a cut-off at the value of the upper limit.

One way to approach this issue is using the so-called Tobit model \citep{Tobin1958}. The model assumes the existence of an unobservable (latent) dependent 
variable $y_i^*$ that replaces the left-hand side of Eq.\,\ref{reg_moderated}\,:\,$y_i^*=X_{ij}\beta_j\,+\,\epsilon_i$. The running index $i$ represents the star, and $j$ is each of the linear regression coefficients with $X_{ij}$ being the matrix of independent variables. The $\epsilon_i$ are assumed to be independent and identically distributed following $\mathcal{N}(0, \sigma^2)$. The observable dependent variable $y_i$ (i.e. the measured Li abundance) is then taken to be equal to the latent variable in the absence of an upper limit $U_i$ and equal to the upper limit otherwise
\vskip-1.5em
\begin{eqnarray}
   y_i = \left\{
     \begin{array}{ll}
       y_i^* & \quad \mathrm{if} \quad \bar{\exists}\,U_i \:\: (\textrm{i.e.} \, y_i^* > U_i) \\
       U_i   & \quad \mathrm{if} \quad y_i^*\leq U_i
     \end{array},
   \right.
\end{eqnarray} 

The application of the Tobit model to handle observation-by-observation censoring leads to a log-likelihood function of the form
\vskip-1.5em
\begin{eqnarray}
\mathcal{L} = \sum_{i \,\in\, y_i^* > U_i} ln \left[ \phi\left(\frac{y_i-X_{ij}\beta_j}{\sigma}\right)/\sigma \right] \,+\, 
              \sum_{i \,\in\, y_i^*\leq U_i}ln \left[ \Phi\left(\frac{U_i-X_{ij}\beta_j}{\sigma}\right) \right] \, \,,
\end{eqnarray}

where $\phi(.)$ and $\Phi(.)$ denote the probability and cumulative density functions, respectively, of the the standard normal distribution. The solution to the linear regression is the set of parameters $int.$, $\beta_{1-4}$, offset, that maximizes the log-likelihood function. 

Using the standard simplex algorithm of \cite{Nelder1965} and taking the full sample of 322 stars into account, we obtained the maximum likelihood estimates for the parameters\footnote{We note that the local minimization algorithm is dependent on the starting conditions. We have used the parameters determined from a multivariable regression on the full sample as initial guesses.}.
These are shown in Table\,\ref{Table:CensoredResults}\,. The estimated value for the offset is again different from zero and still larger than the average error bar on Li abundance. 

We estimate the significance of the offset using bootstrap analysis in a similar way as done in Sect.\,\ref{sec:MRA}\,. For the non-censored data, we proceed as before. Because the upper limits do not have such an objective measure of uncertainty, for this subset of the data we sample from a Gaussian distribution centered on the value of the upper limit $U_i$ and with a standard deviation given by the mean of all the uncertainty values (0.07). The fit is repeated 10\,000 times on these generated Li values resulting in 10\,000 maximum likelihood estimates for the parameters. 

The value of the standard deviation of the resulting distribution of the offset values places the observed value of -0.13 at more than 18$\sigma$ from 0. The significance of this result is dependent on the uncertainty associated to the Li abundance measurements, like before, and in this case on the upper limits value too.





\begin{table*}
\caption{The parameters for each coefficient resulting from the censored regression analysis.} 
\label{Table:CensoredResults}
\centering
\begin{tabular}{ccccccc} \hline\hline
 \ \ dataset &  {\it int.} &$\beta_1$ (log(\teff))   & $\beta_2$ ([Fe/H]) & $\beta_3$ (log\,$g$) & $\beta_4$ (log(Age)) & offset \\ 
 \hline
combined sample &  -297.78 & 81.85 & -1.21 & -1.81 & -1.60 & -0.13  \\
combined sample, Jupiter-mass only & -293.42 & 80.77 & -1.21 & -1.88 & -1.65 &-0.18 \\
\hline 
\end{tabular}
\end{table*}

Repeating the procedure of the previous Section, we note that the significance of an offset value that is different from zero drops to about 6$\sigma$ when each error bar (including the uncertainty assumed for the upper limits) is inflated by a factor $\alpha=2$. When $\alpha=5$ this drops to 2.7$\sigma$. 
The results of considering only Jupiter-mass planets are also shown in Table\,\ref{Table:CensoredResults}\,. The offset value of -0.18, slightly larger than for the whole sample, is significant at the 18$\sigma$ level, a value that drops to 8.4$\sigma$ and 2.7$\sigma$ when $\alpha$ are 2 and 5, respectively. Just like for the previous section, this test shows that the typical uncertainty on the Li abundance would have to be underestimated by a factor of 5 for the results to be rendered non-significant.

\section{Discussion}\label{sec:Discussion}

\subsection{The validity of the different working hypotheses}

In the two preceeding sections, we have shown that an offset indicative of Li depletion is recovered between planet hosts and non-host stars. The two methods employed reveal the same tendency but with a different amplitude; still, an absolute agreement was not expected, as we were performing the regression on different data. Since the Li abundance upper limits values are, by nature, very small, and so for both hosts and non-host stars (as they depend mostly on the quality of the spectra), any offset derived between the two populations making heavy use of these values will be by construction artificially small. As a thought experiment, if one takes the spectra of the full dataset and degrades the resolution progressively, the stronger the degradation, the larger the number of Li lines which will be too small to have their abundance measured, and to which only an upper limit can be assigned. In the extreme case of having only upper limits available for our analysis, the offset between the two groups would be by construction arbitrarily small. One could reach a better agreement between both results by considering skewed error bars for the upper limits, representing an increased probability of drawing lower values rather than high. However, we consider this as a very artificial validation procedure for a situation in which the error bars cannot be robustly characterized. 

It is also very important to recall that the upper limits should always be considered with caution. The measurement of the upper limits is subject to a much higher degree of subjectivity than the measurements themselves, as they often require a visual inspection of the spectra and are thus very difficult to automatize. For the two reasons mentioned above, we consider the determination of Li depletion as obtained from Sect.\,\ref{sec:MRA} as more representative of reality.

It is well known that the independent variables used for the regression analysis are correlated \citep[e.g.][]{1994A&A...287..191P}. This issue, termed multicollinearity by statisticians, has been shown not to have an impact on the values obtained from regression. \cite{Goldberger1991} shows how the existence of a correlation between the independent variables is equivalent to a reduction of the sample size and does have an impact on the test of the hypothesis. While we caution the reader on the interpretation of the coefficients \textbf{$\beta_i$} due to the correlation between the terms, we stress that the method used here is fully consistent.

The linear dependence hypothesis employed in this work has already been made by several other authors (see references in Sect.\,\ref{sec:Intro}\,) as a first approximation to the real dependence on stellar parameters, which is probably more complex. A linear dependence of the envelope of Li abundance on \teff  \, is apparent in the Fig.\,2 of DM14; our analysis indicates that this parameter is the one to which the strongest linear dependence is associated. For the other parameters, the exact functional form of the dependence remains more elusive, but several works have suggested a linear correlation, like the log(Age)-Li abundance connection proposed by \cite{2005Sci...309.2189C}. Importantly, for all the other parameters than the Age, the range of values explored is much narrower, and we stress that our working hypothesis is only that the linear hypothesis holds within this limited range.

A more complex model could be attempted, and an exhaustive search of its suitability can be explored using interaction terms for the proposed components. However, such a model would remain far more speculative than the simplest approach we propose here, especially before the limitations of linear model are assessed. This is particularly clear when we take into consideration that the scatter in Li abundances might be of astrophysical origin and thus cannot be used as a direct argument to evaluate the performance of the model. It is patent in the Li abundance scatter measured in old stellar clusters, like M67 \citep{2007A&A...469..163R, 2008A&A...489..677P, 2012A&A...541A.150P}, and to a lesser extent in other clusters, such as NGC\,3960 \citep{2007A&A...475..539P}, Collinder\,261 \citep{2005ESASP.560..867P}, and NGC\,6253 \citep{2010IAUS..268..275R}, where the stars are supposed to have similar stellar parameters. 

It has been argued that a systematic difference of stellar parameters between the planet-host stars and the comparison stars could lead to a measured Li abundance difference between the two groups and be misinterpreted as Li depletion \cite[as proposed by e.g.][]{ramirez_li12}. It is exactly the objective of our approach to consider the effect of stellar parameters' dependence on Li abundance, as it places the two populations on even footing and considers them equally for the definition of a common regression slope. The constant that is associated to the moderator/dummy variable shows that a real difference in offset exists between the two groups when a linear relationship is assumed.

It is interesting to note that the results hold when only Jupiter-mass planets are considered. This is of particular importance because the samples used in this work are expected to be fully characterized in what concerns the presence of these planets, a point which cannot be fully enforced for the lower-mass regime. Moreover, and as expected, the recovered offset for massive planets is slightly larger. A larger Li depletion for massive planets is in line with the standard theory, as their larger and longer-lived disks would have a greater effect on the rotational history of the star \citep{bouvier08}, and the accretion processes are expected to have a larger impact on Li destruction \citep{baraffe10,theado12}. The difference in offsets when considering stars with different planetary populations is, however, small and the need for more data, especially {\it bona fide} measurements, is very clear.

\subsection{The robustness of the method}\label{sec:addtests}

A question that can arise is if the multivariable regression as described in Sect.\,\ref{sec:MRA} can artificially introduce an offset between the two populations. To address this issue we explored the impact on the offset value of contaminating our planet-host sample with comparison stars. Since in the extreme scenario (i.e. when all the planet-hosts are replaced by comparison stars) we will be working only with comparison stars, we will have to use a lower number of stars in the mock samples than in the real ones; their sum has to be smaller or equal than 139, the total number of comparison stars with Li abundance measurements available.\footnote{We note that if we had an infinitly large number of comparison host stars available, we could draw from this pool of stars to create groups with any size, and would thus be able to reproduce the same scenario as observed in what concerns number of stars. As we cannot, we chose to work with a smaller number of stars rather than to use a variable number of stars on the samples or to use a star repeatedly.} In this way we can keep constant the number of stars considered while varying only the number of real planet hosts included in the mock planetary-host sample. We choose then to work with a sample size of 30 for our mock planet-hosts and a sample size of 100 for our mock comparison stars. 

\begin{figure}

\includegraphics[width=9.5cm]{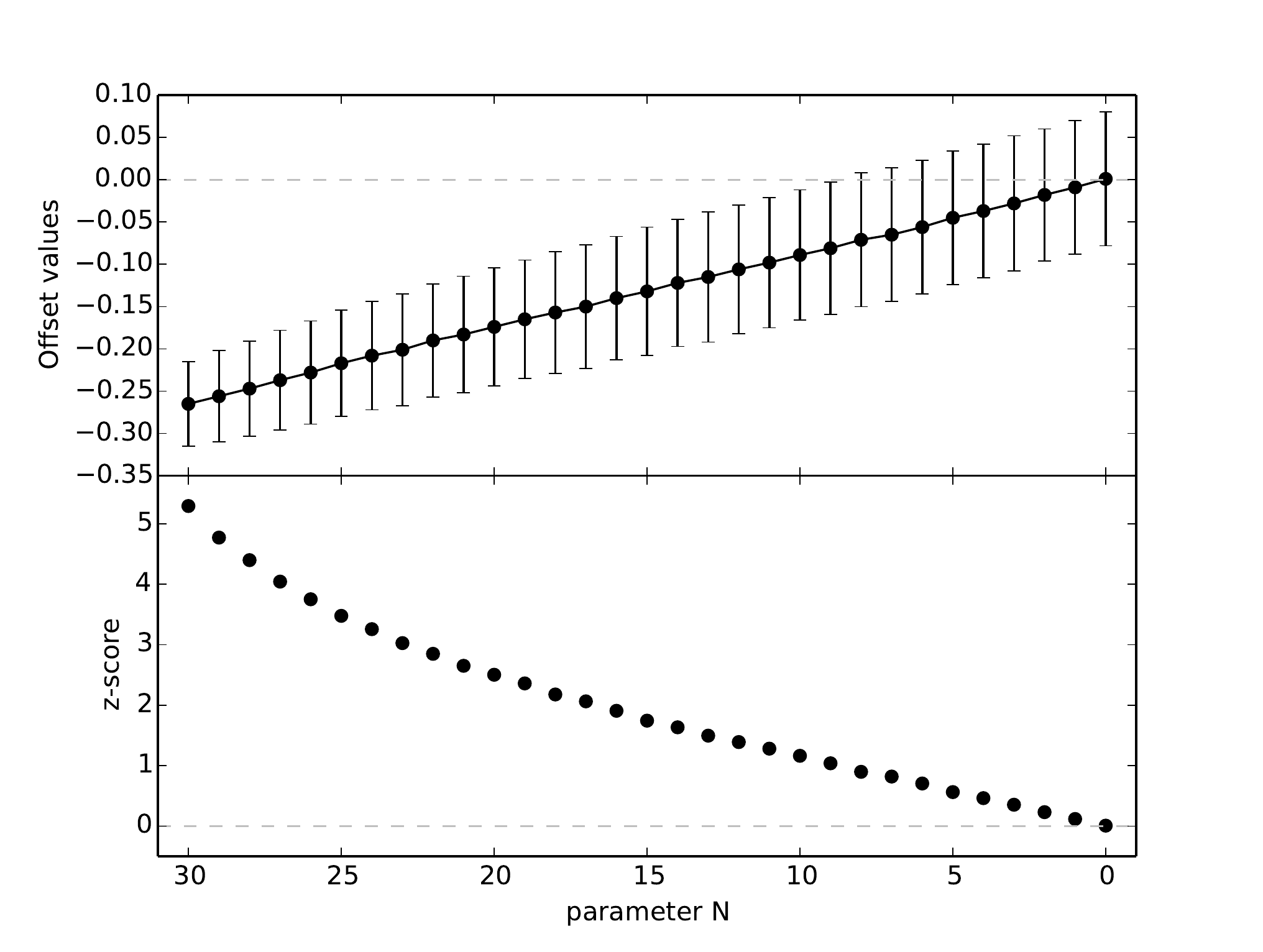}

\caption{The result of poluting a mock planet-host sample with comparison stars following the procedure described in Sect.\ref{sec:addtests}. {\it Upper panel}: Dependence of the average offset and associated dispersion on the number $N$ of planet-host stars in the mock planet-host star population of 30 stars. {\it Lower panel}: Z-score of the offset as calculated relative to zero as a function of the same parameter $N$. The dashed line is a visual aid, representing an offset of zero on the upper panel and a z-score of zero in the lower one.}\label{Fig:add_off}

\end{figure}

To perform this test, we randomly draw $N$ planet-host stars from our sample of 44, to which we add (30-$N$) stars randomly drawn from our sample of comparison stars to form the mock planet-host sample. From the remaining comparison stars, we draw 100 stars to form the mock comparison sample. We note that no star was used twice following this procedure; in other words no star was included in the two groups simultaneously. We applied then the multivariable regression analysis described in Sect.\ref{sec:MRA} following the moderated regression described by Eq.\ref{reg_moderated}. We repeated the procedure 10 000 times and analyzed the distribution of the offset as delivered by the regression, calculating its mean value, dispersion, and corresponding z-score as measured relative to zero. The results of the test are presented in Fig.\,\ref{Fig:add_off}. The value of the offset decreases in an approximately linear fashion as we introduce more and more comparison stars in the mock planet-host sample, showing that the offset indeed stems from the presence of planet-hosts in the sample. The z-score and (thus significance) of the results decreases with decreasing offset absolute value, reaching a value of zero when there are no planet-hosts in the mock-planet population. The take-away lesson from this simple test is that the different properties of the stars in the two groups are at the root of the measured offset.

A different test can be performed by shuffling the tags of planet-hosts and comparison stars. This corresponds to the alternative scenario in which it is the presence of planets around a star that is ultimately unknown (and not its Li abundance that is uncertain). To perform this test we randomly picked 44 stars out of the ensemble of 44 planet-hosts plus 139 comparison stars and attributed them the planet-host tag, leaving the remaining 139 as comparison stars. This test thus assumes the tagging is incorrect but retains the relative proportion of planet-hosts and comparison stars in the sample. We repeated the procedure 10 000 times and calculated, once again, the z-score of the offset value of -0.26 obtained from Sect.\,\ref{sec:MRA} relative to this distribution and the probability associated to it. The offset is at 3.75\,$\sigma$, and the probability associated to it is smaller than 0.1\%. However, one has to bear in mind that the populations created by shuffling in this way contain (a random number of) actual planetary-hosts tagged as comparison stars, of which depletion, when present, will be absorbed by the linear dependence, as discussed in the second paragraph of Sect.\,\ref{sec:MRA}. This effect shifts the zero-point defined by the comparison stars against which the planetary depletion offset is measured. Moreover, this ``polluting'' effect further depends on the relative number of planet-hosts to comparison stars. To illustrate this point, we selected populations of 30 planet-hosts and 100 comparison stars (as described as the starting point for the test of the previous paragraph). While maintaining the total number of stars used constant, we varied the ratio of planet-hosts to comparison stars from 30/100 to 35/95, 40/90, and finally 44/86. While the offset obtained was of -0.26 for all cases, the significance as delivered by shuffling the tags for each of these samples changed from 3.10 to 3.28, 3.35 and 3.42, respectively, showing that the ratio between the number of elements has an impact on the estimated significance. As a consequence, we caution the reader that the result of this test should be seen merely as indicative.

To further test the robustness of our results, we repeated the tests of Sections\,\ref{sec:MRA} and \ref{sec:Censored} for several restricted datasets: by selecting only stars with \teff = {T$_\mathrm{eff, \odot}\pm80K$, \citep[see e.g.][]{israelian09}, stars that host planets with masses larger than 1\,M$_{\mathrm{Jup}}$ and by considering the absolute value of the variables in Eq.\,\ref{reg_moderated}. The offset is always of similar value with a tendency to being slightly larger, and always of similar significance. One has to be careful with the interpretation of such results, and it is not our objective to fine-tune the model to obtain the largest or more significant offset. However, it proves that the results obtained here are a general property of the data, and that the planet hosts do show a measurable Li abundance depletion when compared with non-hosts.
While an offset between the two samples is clearly present, we would like to stress that its most probable absolute value and significance depends strongly on the assumed Li abundance uncertainties. As shown at the end of Sect.\,\ref{sec:MRA} and Sect.\ref{sec:Censored}, if a linear dependence of Li abundance on stellar parameters is assumed, a sizeable increase of the error bars by a factor of 5 or larger is necessary for the significance of the calculated offset to drop below 3-4\,$\sigma$ and, thus, for it to be compatible with zero.

\section{Conclusions}\label{sec:Conclusions}

We have shown that the dataset at hand shows a significant Li depletion for planet hosts if a linear relation between the fundamental stellar parameters log(\teff), [Fe/H], log\,$g$, and log(Age), and Li abundance is assumed. This result is obtained by considering both only stars with Li abundance determinations and the full dataset containing those with upper limits. 

The common working hypothesis of a linear relationship between log(A(Li)) and the different parameters is clearly an oversimplification but is a first step towards a finer understanding of the problem. While not being identical to G14 $\Delta_{p,c}$ (Eq.\,\ref{Gonzalez:Delta}) we note that our Eq.\,\ref{reg_moderated} is equivalent, as the only difference is in the logarithm of the ratios, and these are absorbed in the intercept term.

Through our analysis, the offset between hosts and non-host is recovered in a robust fashion, and in that sense the primary objective of the paper was accomplished, but one has to bear in mind that the result is subject to the limitations inherent to both the model and the data. When applying the model delivered by the analysis done in Sect.\,\ref{sec:MRA} and subtracting it, the standard deviation from the mean of the measured Li abundances (i.e. not including upper limits) decreases from 0.46 to 0.38 dex, which is a small decrease. This residual scatter can be either of physical origin, being rooted on non-modeled physical processes, or stem from an underestimation of the Li uncertainty. We have shown that a very large increase in the Li abundance uncertainty of approximately five times the average error bar is required to reduce the offset z-score to below 3-4\,$\sigma$ and thus make it statistically non-significant. An increase in Li uncertainty by a factor of five would explain the observed residual Li abundance, so we note that it remains a possible option, while explaining the current results. 

Finally, we demonstrated that the offset, as delivered by our method depends on the different nature of the stars in the two samples. We did so by showing that the offset is reduced down to zero if the planet-host stars are replaced by comparison stars in a mock planet-host sample. Importantly, a relative Li depletion effect is also recovered when only stars within a restricted range of stellar parameters are selected, or only planetary hosts with Jupiter-mass planets are considered. 
These tests re-inforce the idea that the conclusions reached here are not the result of a fine-tuned analysis but a general property of the data.


\begin{acknowledgements}
This work was supported by the European Research Council/European Community under the FP7 through Starting Grant agreement number 239953. PF and NCS acknowledge support by  Funda\c{c}\~ao para a Ci\^encia e a Tecnologia (FCT) through Investigador FCT contracts of reference IF/01037/2013 and IF/00169/2012, respectively, and POPH/FSE (EC) by FEDER funding through the program ``Programa Operacional de Factores de Competitividade - COMPETE''. 
EDM, SGS, and VZhA acknowledge the support from the Funda\c{c}\~ao para a Ci\^encia e Tecnologia, FCT (Portugal) in the form of the fellowships SFRH/BPD/76606/2011, SFRH/BPD/47611/2008, and SFRH/BPD/70574/2010 from the FCT (Portugal). GI acknowledges financial support from the Spanish Ministry project MINECO AYA2011-29060.
PF and JF further thank the CrossValidated online community for enlightening discussions on statistical tests. We warmly thank all those who develop the {\it Python} language and its scientific packages and keep them alive and free. We acknowledge the anonymous referee for insightful comments and suggestions that increased the general quality of the paper.

\end{acknowledgements}

\bibliographystyle{aa} 
\bibliography{Mybibliog,extra} 

\end{document}